\documentclass[twocolumn,twoside,slac_two]{revtex4}
\usepackage{graphicx}
\usepackage{fancyhdr}
\begin{document}

\title{Inclusive semileptonic B decays}
\author{Elisabetta Barberio}
\affiliation{School of Physics, University of Melbourne, Parkville, Vic, Australia}

\begin{abstract} The present status of the measurement of 
the semileptonic  B decays
 is reviewed. Emphasis is given on the factors 
affecting experimental errors of the
 Cabibbo-Kobayashi-Masawa matrix element $|V_{cb}|$ and  $|V_{ub}|$. 
\end{abstract}

\maketitle

\thispagestyle{fancy}

\section{ Introduction}

In the framework of the Standard Model, the quark sector is characterised by a rich pattern 
of flavour-changing transitions, described by the Cabibbo-Kobayashi-Maskawa (CKM) matrix.
The quark transitions $B  \to X_c \ell \nu$ and $B  \to X_u \ell  \nu$ provide a way 
for determining these CKM matrix elements.  
This report reviews the experimental measurements used to extract  the 
quark mixing parameters $ |V_{cb}|$  and $|V_{ub}|$ from the inclusive
branching fractions of semileptonic b-hadron decays $Br (B \to X \ell  \nu )$. 
At present,  the values of $|V_{qb}|$ obtained from  inclusive semileptonics decays are more precise than the corresponding exclusive determinations. 

All theoretical approaches  used to extract $|V_{qb}|$  exploit  the fact that the 
mass of the b quark is large compared to the scale $\Lambda_{QCD}$ that 
determines low-energy hadronic physics. The basis for precise calculations is the expansion in 
powers of  $\Lambda_{QCD}/m_b$ , where effective-field-theory methods are used to separate 
non-perturbative from perturbative contributions.

Several studies have shown  that the spectator model decay rate is the leading term in a well-defined 
expansion controlled by  the parameter $\Lambda_{QCD}/m_b$. 
Non-perturbative corrections to this leading approximation arise only to order
 $1/m_b^2$. The key issue in this approach is the ability to separate non-perturbative 
 corrections, that can be expressed as a series in powers of $1/m_b$, and perturbative 
 corrections, expressed in powers of  $\alpha_s$. 
There are different methods \cite{gremm-kap} \cite{falk} \cite{142} \cite{143} \cite{adam1} \cite{g} 
to handle the energy scale $\mu$ used to separate 
long-distance from short-distance physics and they are 
discussed in the theory part of these proceedings.
 The coefficients of the $1/m_b$ power terms are expectation 
values of operators that include non-perturbative physics. 
 
 In this framework, non-perturbative corrections are parameterized by quark masses and matrix elements of higher dimensional operators that at present are poorly known. 
The experimental accuracy already achieved, and the one expected 
from the large data sets recorded by the B-factories, make the ensuing 
 theory uncertainty a major limiting
factor. The extraction of the non-perturbative parameters, describing the heavy 
quark masses, kinetic energy of the b quark and the $1/m_b^3$ corrections, directly from the data 
has therefore become a key issue.  Recently, a lot of progress has been
made on the use of the moments of energy and mass spectra in $B \to X_c \ell  \bar{\nu}$ and 
 $B \to X_s \gamma$ for performing these determinations. 

\section{$ B \to X_c \ell  \nu $}

The determination of  $|V_{cb}|$ from inclusive decays currently has  
an error  of $2\%$, dominated by the knowledge of higher order perturbative and 
non-perturbative corrections. This method is based on the measurement of the 
total semileptonic decay rate, together with the leptonic energy and the hadronic 
invariant mass spectra of inclusive semileptonic decays and $B \to s \gamma$ transitions. 

The non-calculable non-perturbative quantities are parametrised in terms
 of expectation values of hadronic matrix elements, which can be related to 
the shape (moments) of inclusive decay spectra. 
The shape of the lepton energy spectrum and of the hadronic mass spectrum 
provides constraints on the heavy quark expansion based on local Operator
Product Expansion (OPE) that describes the properties of the $Br(B \to X_c \ell  \nu )$ transitions.

So far, measurements of the hadronic mass distribution and the leptonic spectrum have been made by 
Babar, Belle, Cleo and DELPHI, while CDF provide only the measurement of the hadronic mass spectum
 \cite{belle} \cite{bb} \cite{bbl} \cite{battaglia} \cite{CDF} \cite{chen2001}.   

The measured hadronic mass distribution and lepton energy spectrum are affected by detector 
resolution, accessible phase space, radiative corrections, etc. It is particularly important 
for reducing both theoretical and experimental uncertainties to have the largest accessible 
phase space in which to measure the decay spectra. Each of the respective experiments has 
focused on lowering the lepton energy cut-off values.

All experiments performed measurements of the hadronic invariant mass spectrum $M^2_X$. 
 Babar has measured up to the second moment of this distribution, DELPHI up to the third. 
  In order to compare with theoretical predictions, the moments are measured with a well 
defined cut on the lepton momentum in the B rest frame.

Belle\cite{belle}, Babar\cite{bb} and CLEO\cite{newmxmom} explored the moments of the
hadronic mass spectrum, as a function of the lepton momentum cuts. CLEO performs 
a fit for the contributions of signal and backgrounds to the full three-dimensional
 differential decay rate 
distribution as a function of the reconstructed quantities $q^2$, $M_X^2$, $\cos{\theta_{\ell}}$. 
 Belle\cite{belle} and Babar\cite{bb}  use a sample where one of the the
 B-mesons is fully 
reconstructed and the signal side is tagged by a well identified lepton. In this case 
the main sources of systematic errors are the uncertainties related the detector and 
signal modelling, and to reconstruction.

\begin{figure}[b]
\begin{center}
\includegraphics[height=48mm]{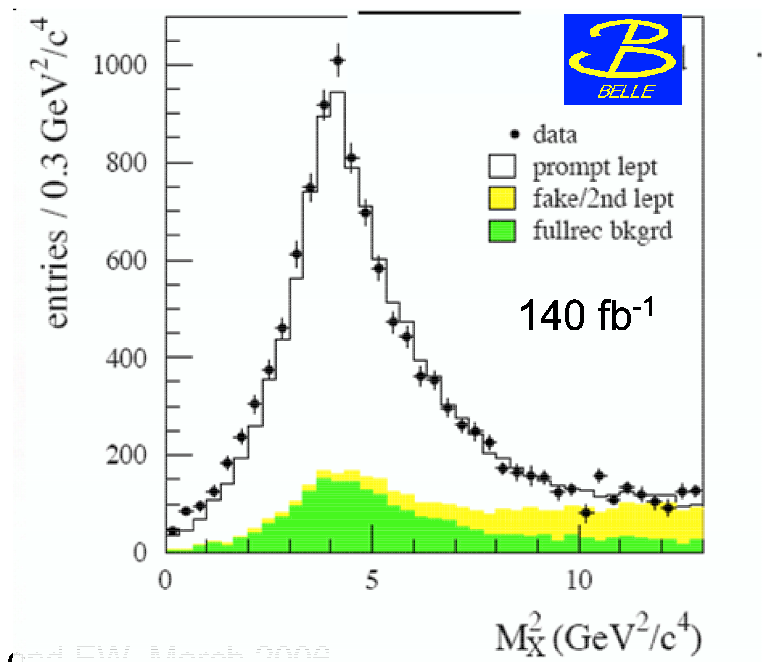}
\caption{Belle hadronic invariant mass spectrum}
\label{fig:hadron}
\end{center}
\end{figure}

The hadronic mass spectrum in $B \to X_c \ell  \nu$ decays can be split into three contributions 
corresponding to $\rm D$, $\rm D^*$ , and $\rm D^{**}$, where  $\rm D^{**}$ 
stands for any neutral charmed state, resonant or not, other than $\rm D$ and $\rm D^*$.
Belle reconstructs the full hadronic mass spectrum and derives  $M_X^2$ 
with an unfolding technique.  Babar extract the moments from the measured distributions
 using a calibration curve derived from Monte Carlo data.  
Employing this method Babar reaches a minimum momentum for the electron 
in the B meson rest frame of 0.9 GeV, while Belle reaches 0.7 GeV, Figure \ref{fig:hadron}.

DELPHI follow a different approach in extracting the moments. The $\rm D$ and $\rm D^*$ 
component of the mass spectrum is well known, hence DELPHI measures the invariant
 mass distribution of the $\rm D^{**}$ component only and fix
the $\rm D$ and $\rm D^*$ components.  In fact,  the  $\rm D^{**}$  component is not well known.
For this reason DELPHI measures the first moment with respect to the spin averaged mass
of $\rm D$ and $\rm D^*$.  At LEP b-quarks were created with an energy of approximately 
30 GeV. The large boost allowed DELPHI  to measure the hadronic mass moments without a cut on
 the lepton energy \cite{battaglia}.

The shape of the lepton spectrum provides further constraints on the OPE. Even if these measurements are less sensitive to OPE parameters they are considerably more precise experimentally.
Moments of the lepton momentum with a cut $p_{\ell}\ge 1.0$ GeV/c
 have been measured by the CLEO collaboration \cite{prd2003}. Babar \cite {bbl}
 extract up to the third moment of this distribution,
using a low momentum cut of  $p_{\ell} \ge 0.6$ GeV/c. Both Babar and CLEO use dilepton samples. 

\begin{figure}
\begin{center}
\includegraphics[height=40mm]{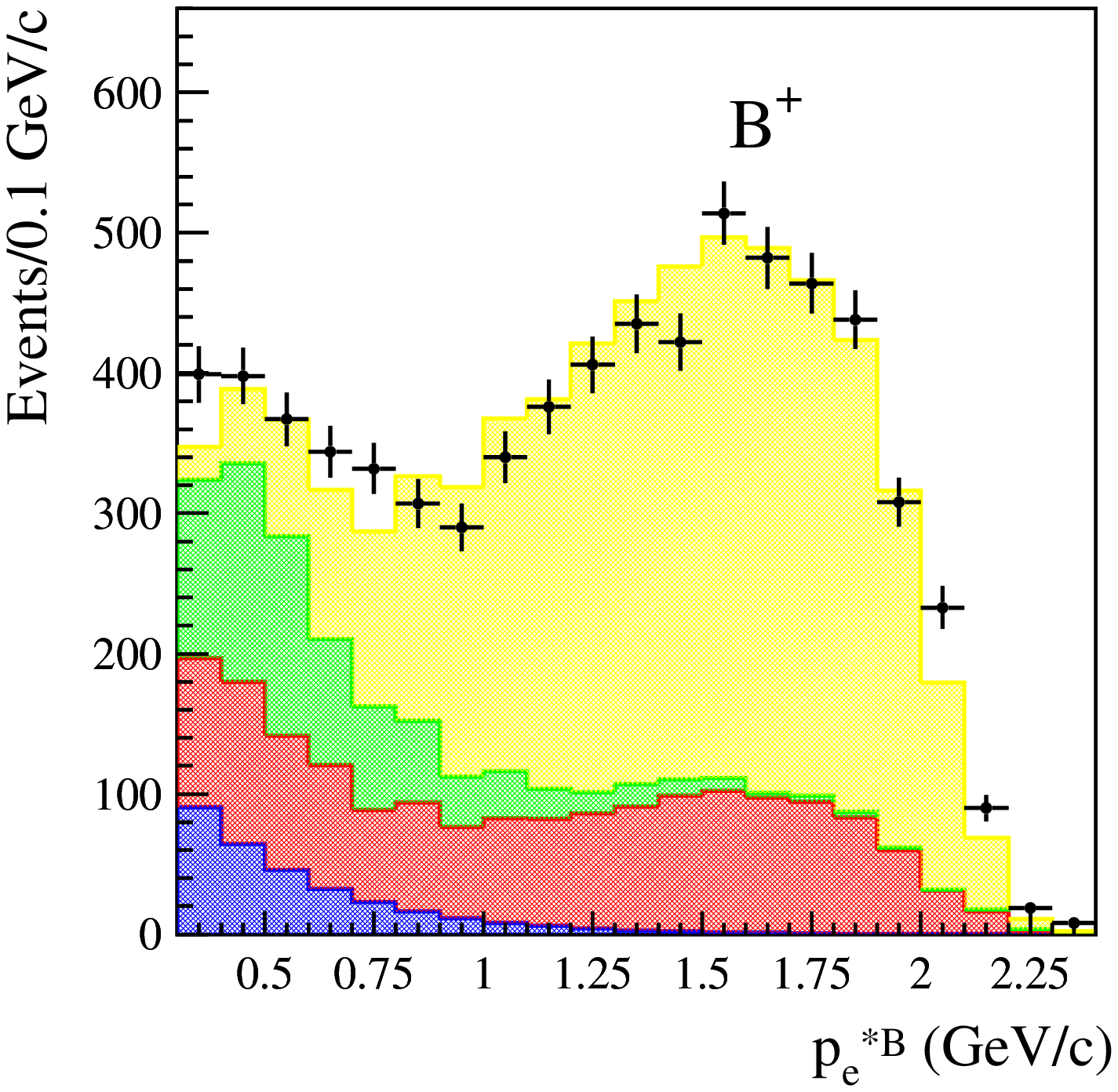}
\includegraphics[height=40mm]{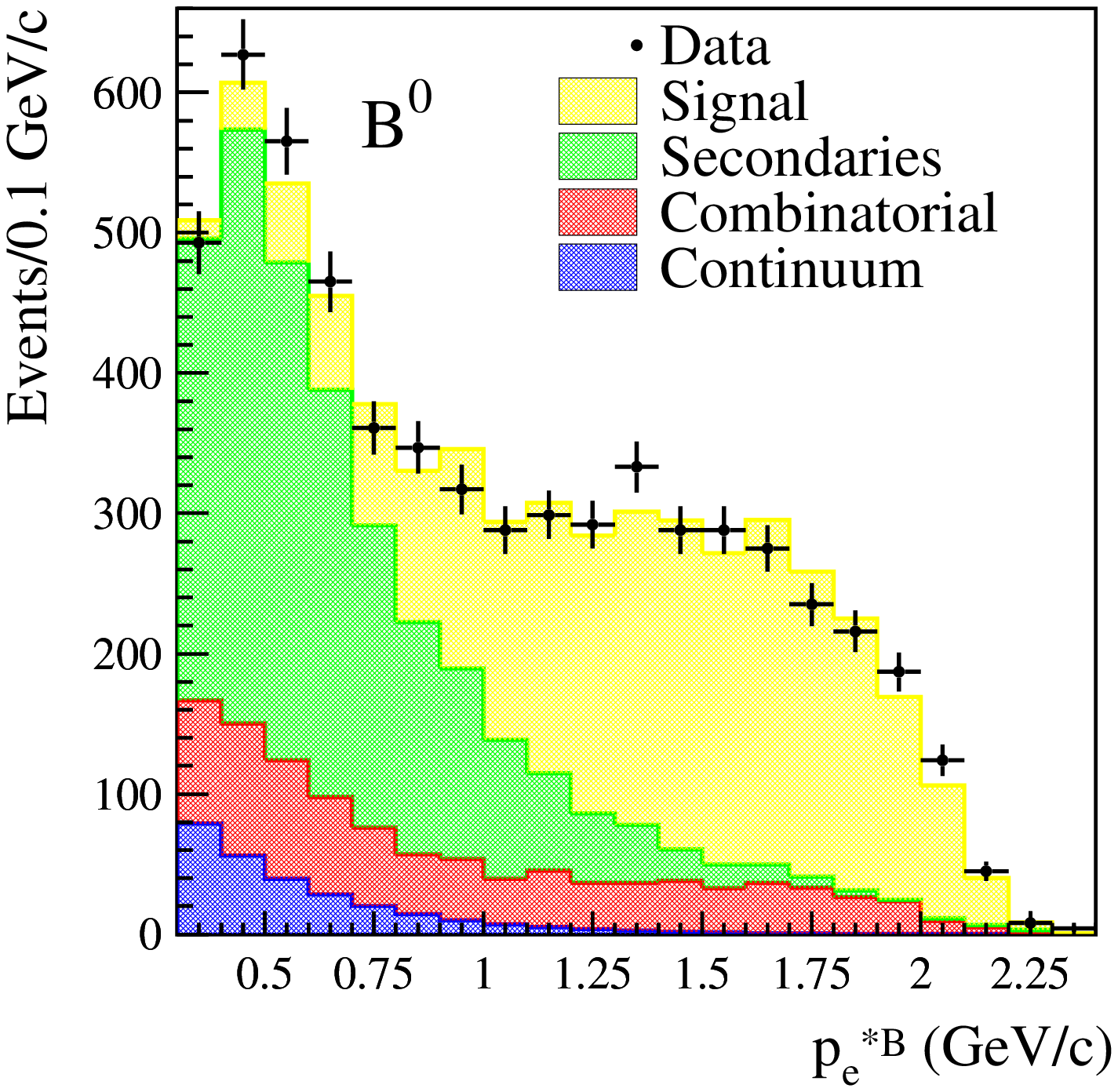}
\caption{Belle:electron energy spectra for $B^0$ and $B^+$.}
\label{fig:lepton}
\end{center}
\end{figure}

Belle \cite{belle} determines the first, second and third moments of the electron energy spectrum 
with a minimum lepton momentum cut ranging between 0.4 and 1.5 GeV in the B meson rest frame.  
Belle measures, independently, the 
electron energy moments for the semileptonic decays of the
$B^+$ and the $B^0$ mesons \cite{belle}, Figure \ref{fig:lepton}.  
In Belle, events are selected by  fully reconstructing one of the B mesons, produced in pairs from 
$\Upsilon (4S)$, in several hadronic decay modes.  Prompt semileptonic decays
 ($b \rightarrow x \ell  \nu$) of the non-tag side $B$ mesons are separated from cascade charm decays
 ($B \rightarrow X_c \rightarrow X \ell  \nu$), based on the correlation between the flavour of 
the tagged B and the lepton charge.  
 DELPHI calculates up to the third moment for the
 lepton energy spectrum, without an explicit lepton momentum cut \cite{battaglia}. 
 
All the lepton moment measurements are consistent with theory and with the moment of the 
 hadronic and $b \to s \gamma$ spectrum. Hadronic and lepton energy measurements are 
consistent within their errors. When compared with theory there is no sign of inconsistencies.

A global fit to all available  hadron mass, electron energy and photon energy moments in 
the 1S scheme has been performed in \cite{55}, giving:
\begin{equation} |V_{cb} | = (41.4 \pm 0.6 \pm 0.1) \times 10^{-3} \nonumber 
 \end{equation}
\begin{equation} m^{1S}_b = 4.68 \pm 0.03 GeV \nonumber 
 \end{equation}
\begin{equation} \lambda_1^{1S} = 0.27 \pm 0.04 GeV \nonumber 
 \end{equation}

where the first error includes experimental and theoretical uncertainties and the second error on 
$|V_{cb}|$ is due to  the B lifetime. 

A more updated  set of data has been fitted in the kinetic 
scheme, resulting in \cite{fit}: 
\begin{equation}|V_{cb} | = (42.0 \pm 0.2 \pm 0.1) \times 10^{-3} \nonumber 
 \end{equation}
\begin{equation}m^{kin}_ b = 4.59 \pm 0.04 GeV  \nonumber 
 \end{equation}
\begin{equation} \mu^2_{\pi}= 0.406 \pm 0.042 GeV  \nonumber 
 \end{equation}
where the first error includes statistical and theoretical uncertainties and the second error on 
$|V_{cb}|$ is from the estimated 
accuracy of the HQE for the total semileptonic rate. 

In both fits, the value of the b and c quark masses are in good agreement 
with independent determination of the same parameters.

\section{$ B \to X_u \ell \nu $}

The decay rate for $B \to X_u \ell \nu $ is proportional to $|V_{ub}|^2$ and 
$m_b^5$. The theoretical description of inclusive $B \to X_u \ell \nu $ decays
 is based on the Heavy Quark Expansion, as for $B \to X_c \ell \nu $ decays \cite{61,62}, 
which predicts the total decay rate with uncertainties of about  5\%. 

Experimentally, the principal challenge is to separate the signal $B \to X_u \ell \nu$  
decays from the 50 times larger $ B \to X_c \ell \nu $  background. This can be achieved
 by selecting regions of phase space in which this background is highly suppressed. In these 
regions the spectra are affected by the distribution of the b-quark momentum inside the B meson,
 which can be described by a structure or "shape function" (SF) \cite{63,64}, in addition to
 weak annihilation and other non-perturbative effects.
 
 Extrapolation from the limited momentum range near the endpoint to the full spectrum 
is a difficult task. 
The shape function is a universal property of B mesons at leading order  \cite{63,64}.
Sub-leading shape functions \cite{66}-\cite{72} arise at each order in $1/m_b$ and 
differ between semileptonic and radiative B decays.
 Several functional forms for the SF, which generally depend on two parameters related to the mass 
 and kinetic energy of the b-quark, $\Lambda$ or $m_b$ , and $\lambda_1$ or $\pi^2$, have been 
proposed. The values and precise definitions of these parameters depend on the specific 
ansatz for the SF, the mass renormalization scheme, and the renormalization scale chosen. 

In inclusive measurements, the most common kinematic 
variables discussed in the literature, each having their own advantages, are the lepton energy 
($E_e$), 
the hadronic invariant mass ($M_X$),  the leptonic invariant mass squared ($q^2$) and the 
light-cone momentum component $P_+ = E_X - |P_X|$.  In all cases, the experiments need to model 
$B \to X_u \ell \nu $  decays in order to calculate acceptances and efficiencies.

\subsection{Endpoint measurements}

\begin{figure}
\begin{center}
\includegraphics[height=50mm]{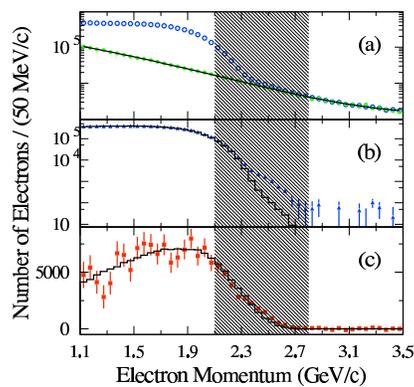}
\caption{BaBar: Electron momentum spectra in the $\Upsilon (4S)$ rest frame:  
(a) on-resonance data (open circles) and scaled off-resonance data (solid circles); 
(b) on-resonance data after non-$\rm B\bar{B}$  background subtraction (triangles)
 compared to simulated $\rm B\bar{B}$  background (histogram); 
 (c) on-resonance data after subtraction of all backgrounds (data points), compared to the 
simulated $B \to X_u \ell \nu$ signal spectrum (histogram).}
\label{fig:l2}
\end{center}
\end{figure}

The first $|V_{ub}|$ measurement performed with this technique was by CLEO. 
In the rest frame of the B meson, the kinematic endpoint of the electron spectrum is about 
2.3 GeV/c 
for the dominant $B \to X_c \ell \nu $ decays and about 2.6 GeV/c for 
$B \to X_u \ell \nu$ decays. The primary challenge in reducing the lepton momentum 
cut in the endpoint method is controlling the $B \to X_c \ell \nu$
 background at the required precision.
The spectrum above 2.3 GeV/c is dominated by electrons from 
$B \to X_u \ell \nu $ transitions, and this allows for a relatively precise measurement,
 largely free from $\rm B\bar{B}$ background, in a 300 MeV/c interval that covers approximately 
$10\%$ 
of the total electron spectrum for charmless semileptonic B decays. Figure \ref{fig:l2}
 shows the Babar lepton energy spectrum distribution for the signal and background.
Belle extracts the 
$B \to X_u \ell \nu$ signal  in the momentum region 1.9 $-$ 2.6 GeV/c,  
Babar covers 2  $-$ 2.6 GeV/c and CLEO 2.3  $-$ 2.6 GeV/c \cite{l1,l2,l3}. 

 \subsection{$M_X$,  $q^2$ and $P_+$ }

As described earlier, this method relies on a sample of events where one of the  B mesons from the
 $\Upsilon(4S)$ decay is fully  reconstructed, while the semileptonic decay of the signal side 
B meson,   is identified by the  presence of a high momentum electron or muon. With
 this sample we are able to
 construct all kinematic variables:   the invariant meson mass $M_X$, 
the lepton neutrino mass squared $q^2$, and the hadronic light-cone momentum 
$P_+$, which are the best available discriminators of signal and background in 
inclusive $|V_{ub}|$ analyses.

BABAR selected semileptonic $B \to X_u \ell \nu $ decays using an approach based on simultaneous requirements for the electron energy, $E_e$, and the invariant mass squared of the $e \nu$ pair, 
$q^2$ \cite{7}. The neutrino 4-momentum is reconstructed from the visible 4-momentum and 
knowledge of the $e^+e^-$ initial state. The dominant charm background is suppressed by 
selecting a region of the  $q^2$ -$E_e$  phase space where correctly reconstructed 
$B \to X_c \ell \nu $ events are kinematically excluded.

 Theoretical studies indicate that it is possible to reduce the theoretical error on the 
extrapolation 
by applying simultaneous cuts on $M_X$and $q^2$ in inclusive $B \to X_u \ell \nu$  decays \cite{l7}.
 In fact, while the $M_X$ distribution has a large usable fraction of events, of the order of $70\%$, 
but depends on the shape function describing the Fermi motion of the b quark inside the B meson, 
the $q^2$ distribution is less sensitive to non-perturbative effects and less dependent on the 
calculation. Unfortunately,  only a small fraction of events (about $20\%$) are usable with a 
pure $q^2$ selection.
The study in \cite{l7} shows that combined cuts on  $M_X$ and $q^2$ mitigate 
the drawbacks of the two methods 
while retaining good statistical and systematic sensitivities. Babar performed a 
measurement of the $|V_{ub}|$  CKM matrix element, on the  fully reconstructed B sample, 
by using the combined information of the $M_X$-$q^2$ distribution to discriminate 
signal and background and to minimize the theoretical uncertainties \cite{b8}.  

\begin{figure}
\begin{center}
\includegraphics[height=40mm]{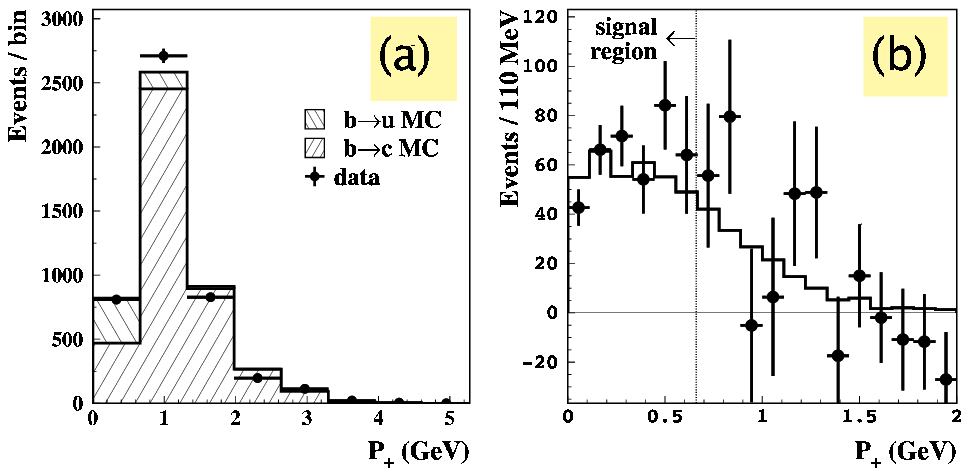}
\caption{The electron momentum spectrum in the $\Upsilon (4S)$ restframe:
 (a) $P_+$ distribution for the selected events. 
 (d) $P_+$ distribution after subtracting the $B \to X_c \ell \nu $ contribution. }
\label{fig:l1}
\end{center}
\end{figure}

Belle  has measured partial rates with cuts on $M_X$, $q^2$ , $E_e$, but the highlight is the 
first measurement of the light-cone momentum $P_+$  \cite{pp}.  Belle measures the partial 
branching fraction in  the kinematic region: $P_+ <$0.66 GeV/c, Figure \ref{fig:l1}. 
 The direct measurement of $P_+$  is quite important as the shape function becomes 
influential when  $P_+$ is not large compared to $\Lambda_{QCD}$ .

\subsection{Reduced Model Dependence}

Babar used  two new techniques to extract $|V_{ub}|$ from inclusive semileptonic B  
decays where the uncertainties due to  $m_b$ and the modeling of the Fermi motion of 
the b quark inside the B meson are significantly reduced \cite{mi}. 
 
\begin{figure}[h]
\begin{center}
\includegraphics[height=40mm]{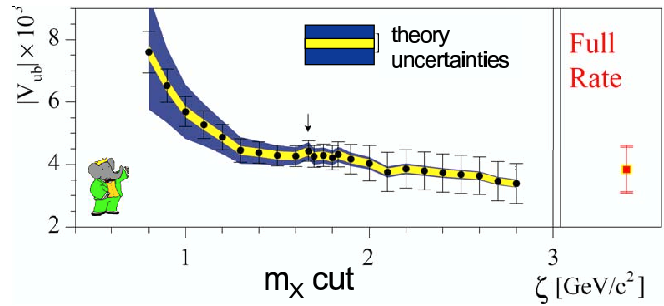}
\caption{$|V_{ub}|$ as a function of  $\xi$ with LLR method (left) 
and for the determination with the full rate measurement 
(right). The errors are statistical only. The shaded area illustrates the theoretical 
uncertainty; the inner 
light shaded (yellow) area indicates the perturbative part of 
the uncertainty. The arrow indicates $\xi$ =1.67GeV/$c^2$.}
\label{fig:hadron2}
\end{center}
\end{figure}

Leibovich, Low, and Rothstein (LLR) presented a prescription to extract 
$|V_{ub}|$  with reduced model dependence from either the lepton energy or the hadronic mass 
$M_X$ \cite{llr}.
A technique to utilise weight functions had been proposed previously by Neubert \cite{n4} 
The calculations 
of LLR are accurate up to corrections of order $\alpha^2_s$ and $(\Lambda m_B / (\xi  m_b ))^2$, 
where $\xi$ is the experimental maximum hadronic mass up to which the  
$ B \to X_u \ell \nu $ decay rate is determined and $\Lambda \sim \Lambda{\rm{QCD}}$. 
This method combines the hadronic mass spectrum, integrated below $\xi$  , with the 
high-energy end of the measured differential $B \to s \gamma $ photon energy spectrum 
via the calculations of LLR. 
An alternative method \cite{al} to reduce the model dependence is to measure the  
$B \to X_u \ell \nu$ rate over the entire $M_X$ spectrum. Since no extrapolation 
is necessary to obtain the full rate, systematic uncertainties 
from $m_b$ and Fermi motion are considerably reduced. Perturbative corrections
 are known to order $\alpha^2_s$. Babar extracts the 
 $B \to X_u \ell \nu $ rate from the hadronic mass spectrum up to 
$\xi$  = 2.5 GeV/c2 which corresponds to about 96\% of the 
simulated hadronic mass spectrum, Figure \ref{fig:hadron2}.

Using the weighting technique with the photon energy spectrum in $B \to s \gamma $ in Ref. \cite{13} 
the hadronic mass spectrum up to a value of $\xi$= 1.67GeV/c2, Babar finds 
$ |V_ {ub} |/|V_{ts}| =0.107\pm 0.009_{stat}\pm 0.006_{syst} \pm0.007_{theo}$.  
Assuming the CKM  matrix is unitary gives 
$|V_{ts} |=|V_{cb}| \times (1\pm O(1\%)) $ and taking $|V_{cb}|$ from \cite{fit}  one gets:
$ |V_ {ub} |=(4.43 \pm 0.38 \pm 0.25\pm ±0.29)\times 10^{-3}$, 
where the first error is the statistical uncertainty, the second is systematic, 
the third theoretical. Babar also determines 
$ |V_ {ub}|$ from  the full $M_X$ spectrum, i.e., up to a value of $\xi$=2.5GeV/$c^2$, and find 
$ |V_ {ub}|=(3.84\pm 0.70_{stat} \pm 0.30_{sys} \pm 0.10_{theo})\times 10^{-3}.$

\subsection{HQE parameters and shape function input }

\begin{figure}
\begin{center}
\includegraphics[height=51mm]{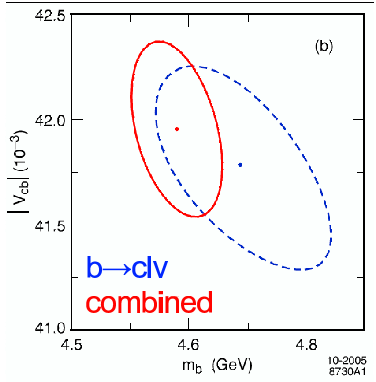}
\caption{Kinetic mass fit from which the heavy quark parameters needed in 
calculating $ B \to X_u \ell \nu $ partial rates are derived. 
The dotted curve shows the fit using information form $ B \to X_c \ell \nu$ 
only, the red line includes information from
$B \to s \gamma $. }
\label{fig:hadron1}
\end{center}
\end{figure}

The global fits to $ B \to X_c \ell \nu$ and $B \to s \gamma $ 
moments in ref \cite{fit} discussed earlier 
provide input values for the heavy quark parameters needed in 
calculating $ B \to X_u \ell \nu $ partial rates and to 
constrain the first and second moments of the 
shape function. Additional information on the leading shape 
function and HQE parameters is obtained from the photon 
energy spectrum in $B \to s \gamma $ decays. The results of the global fit 
are shown in Figure \ref{fig:hadron1}. 

\subsection{$|V_{ub}|$ extraction}

\begin{figure}
\begin{center}
\includegraphics[height=50mm]{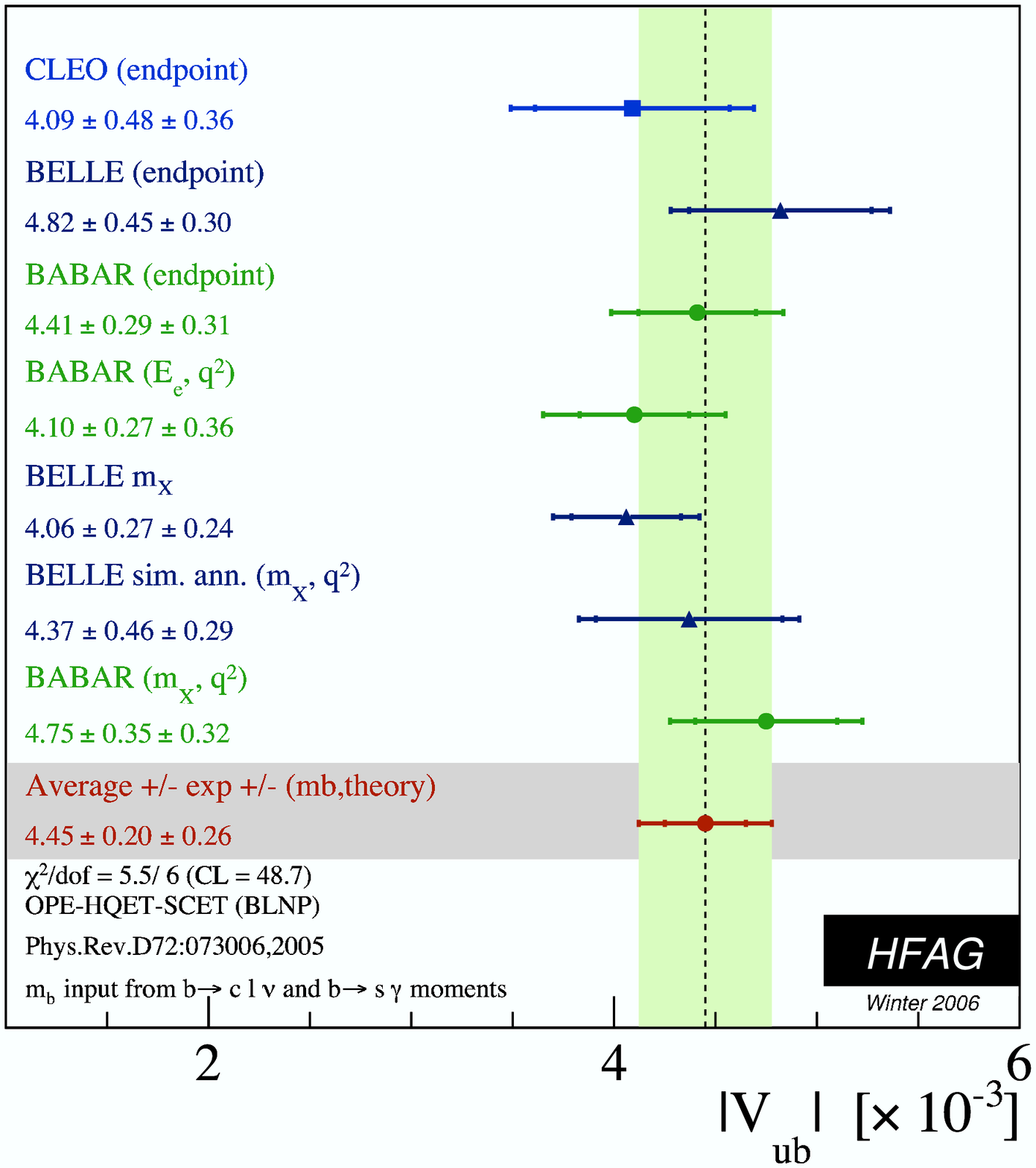}
\caption{ The world average for $|V_{ub}|$}
\label{fig:lepton1}
\end{center}
\end{figure}

The CKM parameter $|V_{ub}|$ is obtained directly from the partial branching fraction using 
$$|V_{ub}|^2  = \Delta B(B \to X_u \ell \nu )/(R \tau_B),$$
 where $\tau_B$  is the average B lifetime. R is the theoretical prediction of the 
partial rate and is calculated for a given signal region by using an inclusive
 $ B \to X_u \ell \nu $ decay 
generator. The extracted values of 
$|V_{ub}|$ are given in Figure \ref{fig:lepton1}.

The world average  is determined by HFAG. We have chosen the value extracted using the BLNP
 \cite{blnp} theoretical framework as the world average of the CKM parameter $ |V_{ub}|$. 
The value of $m_b$ used in the world average is from the global fit in the Kinetic scheme 
to measurements of 
$B \to X_c \ell \nu $ and $B \to s \gamma $ decays.

HFAG also extracts $ |V_{ub}|$ using the Dressed Gluon Exponentiation \cite{dge} framework, a 
 recent new addition to the phenomenology landscape of inclusive B-meson decays. In this
 framework the on-shell b-quark calculation, converted into hadronic variables, 
can be directly used as an approximation to the meson decay spectrum, without 
need of a leading-power non-perturbative function i.e no shape function. 
The on-shell mass of the b-quark within the B-meson is required as an input 
($m_b$). Theoretical uncertainties are assessed by varying $m_b$, the strong coupling 
constant ($\alpha_s$), the number of light fermion flavours, and the method and scale 
of the matching scheme intrinsic to the approach. As the DGE error is not yet fully calculated
 (e.g. the weak anihilation contribution is missing), we prefer not to quote this method as
 the principal one. It must be noted that the value of $|V_{ub}|$  obtained within the DGE
 framework agrees very well with the BLNP determination.

At present, as indicated by the average given above, the 
uncertainty on $|V_{ub}|$ is at the $7\%$ level. The uncertainty on $m_b$ used here is 40 MeV, 
contributing an uncertainty of $4.5\%$ on $|V_{ub}|$.

\section{Conclusions}
 
 The global fits to the semileptonic and radiative decay spectral moments  are used to extract
 the CKM matrix element $|V_{cb}|$  with $2\%$ precision dominated by HQE measurements. 
 
Various methods to extract $|V_{ub}|$ are used, giving a $7\%$ 
precision in the BLNP scheme, dominated by HQE parameters.

\end{document}